\documentclass[conference]{IEEEtran}
\IEEEoverridecommandlockouts

\usepackage{cite}
\usepackage{amsmath,amssymb,amsfonts, bm}
\usepackage{lipsum}
\usepackage{algorithmic}
\usepackage{graphicx}
\usepackage{textcomp}
\usepackage{xcolor}
\def\BibTeX{{\rm B\kern-.05em{\sc i\kern-.025em b}\kern-.08em
    T\kern-.1667em\lower.7ex\hbox{E}\kern-.125emX}}
\begin{document}

\title{Data-Driven Model Identification of Unbalanced Induction Motor Dynamics and Forces using SINDYc\\
\thanks{This work was partially funded by the Flemish Funding Agency (FWO) with grant number 1S02523N. The remainder was supported by the joint DFG/FWF Collaborative Research Centre CREATOR (DFG: Project-ID 492661287/TRR 361; FWF: 10.55776/F90) at TU Darmstadt, TU Graz, and JKU Linz.}
}

\author{\IEEEauthorblockN{1\textsuperscript{st} Emma Vancayseele}
\IEEEauthorblockA{\textit{Department of Physics and Astronomy} \\
\textit{KU Leuven}\\
Kortrijk, Belgium \\
emma.vancayseele@student.kuleuven.be}
\and
\IEEEauthorblockN{2\textsuperscript{nd} Philip Desenfans}
\IEEEauthorblockA{\textit{Department of Electrical Engineering} \\
\textit{Flanders Make @ KU Leuven} \\
\textit{KU Leuven}\\
Bruges, Belgium \\
philip.desenfans@kuleuven.be}
\and
\IEEEauthorblockN{3\textsuperscript{rd} Zifeng Gong}
\IEEEauthorblockA{\textit{Department of Electrical Engineering} \\
\textit{Flanders Make @ KU Leuven} \\
\textit{KU Leuven}\\
Bruges, Belgium \\
zifeng.gong@kuleuven.be}
\and
\IEEEauthorblockN{4\textsuperscript{th} Dries Vanoost}
\IEEEauthorblockA{\textit{Department of Electrical Engineering} \\
\textit{Flanders Make @ KU Leuven} \\
\textit{KU Leuven}\\
Bruges, Belgium \\
dries.vanoost@kuleuven.be}
\and
\IEEEauthorblockN{5\textsuperscript{th} Herbert De Gersem}
\IEEEauthorblockA{\textit{Electromagnetic Field Theory} \\
\textit{TU Darmstadt}\\
Darmstadt, Germany \\
herbert.degersem@tu-darmstadt.de}
\and
\IEEEauthorblockN{6\textsuperscript{th} Davy Pissoort}
\IEEEauthorblockA{\textit{Department of Electrical Engineering} \\
\textit{Flanders Make @ KU Leuven} \\
\textit{KU Leuven}\\
Bruges, Belgium \\
davy.pissoort@kuleuven.be}
}

\maketitle

\begin{abstract}
 This paper identifies the stator curents, torque and unbalanced magnetic pull (UMP) of an unbalanced induction motor by the System Identification of Nonlinear Dynamics with Control (SINDYc) method from time-series data of measurable quantities. The SINDYc model has been trained on data coming from a nonlinear magnetic equivalent circuit model for three rotor eccentricity configurations. When evaluating the SINDYc model for static eccentricity, torques and UMPs with excellent accuracies, i.e., 8.8 mNm and 4.87 N of mean absolute error, respectively, are found. When compared with a reference torque equation, this amounts to a 65\% error reduction. For dynamic eccentricity, the estimation is more difficult. The SINDYc model is fast enough to be embedded in a control procedure.
\end{abstract}


\section{Introduction}
Induction motors (IMs) are widely used because of their robust construction, low cost, ease of maintenance, and high efficiency. However, machine unbalances inevitably occur in the form of electrical, magnetic, or mechanical asymmetries, e.g. due to manufacturing tolerances \cite{salah2019review}. For example, the air gap length is significantly affected by the positioning and straightness tolerances on the rotor axle, causing inherent static or dynamic radial rotor eccentricity. The eccentricity is called static when there is a time-invariant radial displacement between the rotor geometrical center and the stator geometrical center. Alternatively, it is called dynamic when the rotor center revolves around the stator center in a circular motion, e.g. due to shaft bending. Although are small initially, they may be exacerbated by faults. Importantly, the eccentric rotor position distorts the magnetic field which produces an undesired radial magnetic force, called the unbalanced magnetic pull (UMP)~\cite{salah2019review}.

Predictive control of IMs is advantageous for applications where performance, efficiency, and dynamic response are critical. Traditionally, physics-based models such as multiple coupled circuits (MCCs) are used in state-space form to project the motor state forward in time \cite{geyer2016model}. However, these models generally assume magnetic and electrical symmetry, neglecting unbalance. Even when fault features are introduced, an identification of the unbalance is required to appropriately select and parametrise the model, as shown in ,e.g., \cite{raj2021statespace}. Instead, data-driven models leverage the availability of measurable data to construct dynamical models which directly utilise measurables, rendering them suitable for control applications of unbalanced drives. 

The sparse identification of nonlinear dynamics (SINDy) algorithm is a data-driven system identification method which provides nonlinear, sparse, and interpretable models. It was first proposed in \cite{brunton2016discovering} and extended to control problems as SINDYc in \cite{brunton2016sparse}. Although applications have originated in the fields of thermodynamics and fluid dynamics \cite{abdullah2023data}, electric machines have been tackled recently. In \cite{ayankoso2023time}, various data-driven methods are compared for the identification of a geared DC-motor. SINDYc was found highly favourable in terms of accuracy, computational efficiency, and interpretability. The method has been explored for brushless DC-motors \cite{dasanayake2024motor} and multiphase IMs \cite{abu2023data}.

In this work, SINDYc is applied specifically to infer the dynamics of three-phase IMs and the electromagnetic forces, i.e. the electromagnetic torque and the UMP. As the UMP aggravates the rotor eccentricity and stresses the bearings, its estimation from measurables is highly valuable.

\section{Theory}
In this section, SINDYc is explained in general. The variables used belong to $\mathbb{R}$ unless specified otherwise.
\subsection{Identification of Nonlinear Dynamical Systems}\label{subsec:theory_sindyc}
The IM is a controlled nonlinear dynamical system, which is described as:
\begin{equation}\label{eq:nonlinear_system}
    \frac{d}{dt}\bm{x}=\bm{f}(\bm{x}, \bm{u}),
\end{equation}
where $\bm{x} \in \mathbb{R}^{m \times 1}$ is the state vector, $\bm{u} \in \mathbb{R}^{o \times 1}$ the input vector, and $\bm{f}: \mathbb{R}^{m \times 1} \times \mathbb{R}^{o \times 1} \rightarrow \mathbb{R}^{m \times 1}$ the system dynamics. In practice, this system is observed in discrete time as:
\begin{equation}
    \bm{x}_{n+1}=\bm{F}(\bm{x}_{n},\bm{u}_{n}),
\end{equation}
which relates to (\ref{eq:nonlinear_system}) through:
\begin{equation}
    \bm{F}(\bm{x}_{n},\bm{u}_{n})=\bm{x}_{n}+\int^{(n+1)\delta}_{n\delta}\bm{f}(\bm{x}(\tau),\bm{u}(\tau))d\tau,
\end{equation}
with $\bm{x}_n$ shorthand for $\bm{x}(n\delta)$ where $\delta \in \mathbb{R}_{+}$ is the time step, and $\bm{F}$ the time-discrete system dynamics.

The system is observed for several $h$-length snapshots of $\bm{x}$ and $\bm{u}$ as:
\begin{align}
    \bm{X}=\begin{bmatrix}
        \bm{x}_{n} \quad \bm{x}_{n+1} \cdots \bm{x}_{n+h-1}
    \end{bmatrix},\\
    \bm{X}^{+}=\begin{bmatrix}
        \bm{x}_{n+1} \quad \bm{x}_{n+2} \cdots \bm{x}_{n+h}
    \end{bmatrix},\\
    \bm{\Upsilon}=\begin{bmatrix}
        \bm{u}_{n} \quad \bm{u}_{n+1} \cdots \bm{u}_{n+h-1}
    \end{bmatrix},
\end{align}
where, e.g., $\bm{X} \in \mathbb{R}^{m \times h}$. To identify the dynamical system (\ref{eq:nonlinear_system}), the time derivatives of $\bm{x}$ are needed. Although direct measurement of the state derivatives is possible for some systems, this is not the case for the IM. Hence, the derivatives are approximated numerically as:
\begin{equation}
    \frac{d}{dt}\bm{X}\approx\bm{D}(\bm{X},\bm{X}^{+}),
\end{equation}
where $\bm{D}$ represents a temporal discretisation method, i.e. second order central differencing in this work.

A library of candidate functions $\bm{\Theta}:\mathbb{R}^{m\times h} \times \mathbb{R}^{o \times h} \rightarrow \mathbb{R}^{p \times h}$ is defined, e.g., as:
\begin{equation}\label{eq:candidate_library}
    \bm{\Theta}(\bm{X},\bm{\Upsilon})=
    \begin{bmatrix}
    \bm{1}\\\bm{X}\\\bm{X}^{2}\\\bm{\Upsilon}\\\bm{\Upsilon}^{2}\\\bm{X}\otimes\bm{\Upsilon}\\\bm{X}^{2}\otimes\bm{\Upsilon}\\ \vdots
    \end{bmatrix},
\end{equation}
which may comprise polynomial, trigonometric, radial-basis, or other functions of $\bm{X}$, $\bm{\Upsilon}$ and their tensor products, denoted by $\otimes$. A weighting matrix $\bm{\Xi} \in \mathbb{R}^{m \times p}$ is determined such that model complexity and accuracy are balanced, e.g., using the sparse regression algorithm LASSO as:
\begin{equation}\label{eq:weight_matrix_optimisation}
    \bm{\Xi}=\underset{\bm{\Xi}^{*}}{\text{argmin}}\left(\frac{1}{2h}\lVert\frac{d}{dt}\bm{X}-\bm{\Xi}^{*}\bm{\Theta}(\bm{X},\bm{Y})\rVert_{2}^{2}+\alpha\lVert\bm{\Xi}^{*}\rVert_{1}\right),
\end{equation}
where increasing the coefficient $\alpha \in \mathbb{R}_{+}$ results in a sparser solution. When $\bm{\Xi}$ is found, a system of nonlinear equations is obtained that represents the measured system as:
\begin{equation}
    \frac{d}{dt}\hat{\bm{x}}=\bm{\Xi}\bm{\Theta}(\bm{x},\bm{u}),
\end{equation}
where $\frac{d}{dt}\hat{\bm{x}}$ denotes the estimate of the state vector time derivative.

\subsection{Identification of Nonlinear Mappings}\label{subsec:nonlinearmapping}
Instead of identifying a nonlinear dynamical system, sparse linear regression can also be used to identify a mapping $\mathcal{M}: \mathbb{R}^{m \times 1} \times \mathbb{R}^{o \times 1} \rightarrow \mathbb{R}^{q \times 1}$ as:
\begin{equation}
    \bm{y}=\mathcal{M}(\bm{x,u}),
\end{equation}
where $\bm{y}$ is the result of the mapping, named the output vector. Analogously to Subsection $\ref{subsec:theory_sindyc}$, a new library of candidate functions $\bm{\Theta}_{\mathcal{M}}$ is constructed according to (\ref{eq:candidate_library}). Hereafter, a Pareto-optimal weighting matrix $\bm{\Xi}_{\mathcal{M}}$ is found using the snapshots $\bm{Y} \in \mathbb{R}^{q \times h}$, i.e.:
\begin{equation}
    \bm{Y}=\begin{bmatrix}
        \bm{y}_{n} \quad \bm{y}_{n+1} \cdots \bm{y}_{n+h-1}
    \end{bmatrix},
\end{equation}
and $\bm{\Theta}_{\mathcal{M}}(\bm{X, \Upsilon})$ that balances complexity and accuracy according to a sparse optimisation, e.g., (\ref{eq:weight_matrix_optimisation}). This results in a nonlinear mapping based on measurements, which represents $\mathcal{M}$ as:
\begin{equation}
    \hat{\bm{y}}=\bm{\Xi}_{\mathcal{M}}\bm{\Theta}_{\mathcal{M}}(\bm{x,u}),
\end{equation}
where $\hat{\bm{y}}$ denotes the estimate of the output vector.

\section{Methodology}
\subsection{Selection of State and Input Variables}
In this work, SINDYc is used to predict the dynamics of the stator current in a stator reference $dq0$ frame denoted by $\bm{i}^{s}_{dq0} \in \mathbb{R}^{3\times1}$. As a result, the state vector is defined as:
\begin{equation}\label{eq:state_vector_def}
    \bm{x}=\bm{i}^{s}_{dq0}=
    \sqrt{\frac{2}{3}} 
    \begin{bmatrix}
    1 & -\frac{1}{2} & -\frac{1}{2}\\
    0 & \frac{\sqrt3}{2} & -\frac{\sqrt3}{2}\\
    \frac{1}{\sqrt2} & \frac{1}{\sqrt2} & \frac{1}{\sqrt2}
    \end{bmatrix}
    \bm{i}^{s}_{abc},
\end{equation}
where $\bm{i}^{s}_{abc} \in \mathbb{R}^{3\times1}$ denotes the stator currents in the $abc$-phase reference frame.

From theory, it is known that a state-space model of a symmetrical IM can be constructed using $\bm{i}^{s}_{dq0}$ and the stator flux linkages $\bm{\lambda}^{s}_{dq0} \in \mathbb{R}^{3\times1}$ to form the state vector and the stator voltages $\bm{v}^{s}_{dq0} \in \mathbb{R}^{3 \times 1}$ to form the input vector \cite{geyer2016model}. However, using the laws of Ohm and Faraday-Lenz, the stator flux linkage can be expressed in terms of the stator current and stator voltage as \cite{desenfans2024influence}:
\begin{equation}
    \bm{\lambda}^{s}_{dq0}=\int\left(-\bm{R}_{dq0}^{s}\bm{i}^{s}_{dq0}+\bm{v}^{s}_{dq0}\right)dt,
\end{equation}
where $\bm{R}^{s}_{dq0} \in \mathbb{R}^{3\times3}$ is a diagonal matrix of stator resistances. The dependency of the stator current dynamics on the stator flux linkage is reduced to a dependency on the time-integrals of the stator currents and the stator voltages using:
\begin{equation}\label{eq:integrals_definition}
    \bm{I}^{s}_{dq0}=\int\bm{i}^{s}_{dq0}dt, \quad \bm{V}^{s}_{dq0}=\int\bm{v}^{s}_{dq0}dt.
\end{equation}
When accompanied by $\bm{v}^{s}_{dq0}$ and mechanical rotor information, the input vector becomes:
\begin{equation}
    \bm{u} = 
    \begin{bmatrix}
    \bm{v}^{s}_{dq0}\\ \bm{I}^{s}_{dq0}\\ \bm{V}^{s}_{dq0}\\ \gamma^{r}\\ \omega^{r}    
    \end{bmatrix},
\end{equation}
where $\gamma^{r}$ and $\omega^{r}$ are the rotor mechanical angle and the mechanical speed, respectively.

By defining current and voltage integrals as new variables in (\ref{eq:integrals_definition}), estimation of the stator flux linkage and knowledge of the $dq0$ stator resistances are no longer required to construct a predictive model of the stator current. However, its usage requires that the integration of $\bm{i}^{s}_{dq0}$ and $\bm{v}^{s}_{dq0}$ does not diverge over time. Although divergence may occur for arbitrary voltage sequences, this is generally not the case for voltage sequences that create a rotating magnetic field in a static $dq0$ reference frame.

Additionally, two nonlinear mappings are learned using sparse linear regression as described in Subsection~\ref{subsec:nonlinearmapping}, one for the electromagnetic torque $T_{e}$ and one for the unbalanced magnetic pull $\bm{F}_{e} \in \mathbb{R}^{2\times1}$.

Lastly, the inclusion of the stator winding temperature in $\bm{u}$ would be required in practice to accommodate time-varying motor temperatures and their effect on $\bm{R}^{s}_{dq0}$.

\subsection{Data-Generation Model}
\begin{table}[tb]
    \centering
    \caption{Main Induction Motor Drive Parameters}
    \begin{tabular}{|c|c|}
        \hline
         \textbf{Property} & \textbf{Value} \\
         \hline
         Motor name & Cantoni 2SIE 80-2B\\
         \hline
         Nominal output power & $1.1$ kW\\
         \hline
         Nominal line voltage & $400$ V\\
         \hline
         Nominal frequency & $50$ Hz\\
         \hline
         Nominal torque & $3.699$ Nm\\
         \hline
         Nominal rotor speed & $2840$ rpm\\
         \hline
         Pole pair number & $1$\\
         \hline
    \end{tabular}
    \label{tab:im_nameplate}
\end{table}

Data are generated using a nonlinear magnetic equivalent circuit \cite{desenfans2024influence} that models the motor of Table~\ref{tab:im_nameplate} in a wye electrical stator configuration. The model is used to generate data subsets by simulating $5$ seconds of motor run-up with a supply voltage $V$ sampled from $[40, 400]$ $\text{V}_{\text{RMS}}$. The supply frequency $f$ is selected to maintain a constant $V/f$ ratio at $8 \text{ V}_{\text{RMS}}/\text{Hz}$. The supply voltage and frequency are gradually increased for the first $1.5$ s. Thereafter, a load torque is sampled from a distribution scaled by the supply voltage, i.e.  $[0, 3.7\cdot V_{\text{RMS}}/400]$ Nm, and resampled every time the steady state is reached. For (i) a centric rotor, (ii) a rotor with $50\%$ air gap static eccentricity, and (iii) a rotor with $50 \%$ air gap dynamic eccentricity, $25$ subsets are generated each and agglomerated to form three datasets, named datasets A, B, and C. At a time step $\delta=10^{-4}$ s, this amounts to $1.25\cdot10^6$ simulated steps per dataset. Additionally, test datasets are generated for the same three rotor configurations using $1$ subset each for model evaluation.

\begin{figure}[t]
    \centering
    \includegraphics[width=1.0\linewidth]{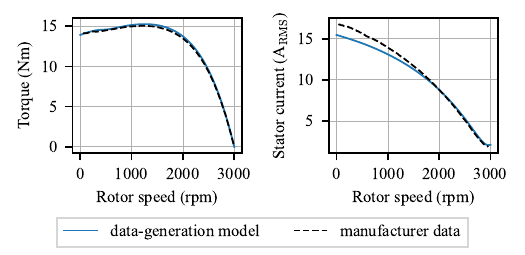}
    \caption{Validation of data-generation model for steady-state motor operation.}
    \label{fig:motorvalidation}
\end{figure}

The evaluation of Fig.~\ref{fig:motorvalidation} shows a close correspondence between the measurement data and the data generation model. The remaining discrepancy of the stator current can be attributed to the negligence of magnetic hysteresis and eddy currents, which are increasingly present at low rotor speeds due to the high rotor frequency.

\section{Results and Discussion}
\subsection{Hyperparameter Optimisation}
When solving for the weighting matrix $\bm{\Xi}$ or $\bm{\Xi}_{\mathcal{M}}$, the selection of the optimiser and its hyperparameters significantly affects model performance. For LASSO, selection of $\alpha$ is required in (\ref{eq:weight_matrix_optimisation}). In addition, the optimisers SR3 and STLSQ are evaluated. SR3 performs the optimisation:
\begin{multline}\label{eq:SR3_optimisation}
    \bm{\Xi}=\underset{\bm{\Xi}^{*},\bm{W}}{\text{argmin}}\biggl(\frac{1}{2}\lVert\frac{d}{dt}\bm{X}-\bm{\Xi}^{*}\bm{\Theta}(\bm{X},\bm{Y})\rVert_{2}^{2}\\+\lambda R(\bm{W})+\frac{1}{2}\nu\lVert\bm{\Xi}^{*}-\bm{W}\rVert_{2}^{2}\biggl),
\end{multline}
which introduces an auxiliary variable $\bm{W} \in \mathbb{R}^{m\times p}$ and regularisation function $R:\mathbb{R}^{m\times p} \rightarrow \mathbb{R}$, e.g., the L1-norm, and uses the hyperparameters $\lambda$ and $\nu$. Lastly, STLSQ performs the optimisation:
\begin{equation}\label{eq:STLSQ_optimisation}
    \bm{\Xi}=\underset{\bm{\Xi}^{*}}{\text{argmin}}\left(\lVert\frac{d}{dt}\bm{X}-\bm{\Xi}^{*}\bm{\Theta}(\bm{X},\bm{Y})\rVert_{2}^{2}+\alpha\lVert\bm{\Xi}^{*}\rVert_{2}^{2}\right),
\end{equation}
similar to LASSO but using a squared L2 regularisation on the weights. 

The selection of candidate functions to form $\bm{\Theta}$ and $\bm{\Theta}_{\mathcal{M}}$ further expands the search space. Here, $5$ libraries are evaluated during model tuning. To simplify their description, we define $\bm{z} = [\bm{x}^{\intercal} \quad (\bm{u}{\backslash\gamma^{r}})^{\intercal}]^{\intercal}$, where $\bm{u}{\backslash\gamma^{r}}$ denotes $\bm{u}$ without $\gamma^{r}$. Library $1$ is created using second-order polynomials of $\bm{z}$ and first-order goniometric terms of $\gamma^{r}$. Library $2$ uses linear terms of $[\bm{i}_{dq0}^{s} \quad \bm{v}_{dq0}^{s} \quad \bm{I}^{s}_{dq0} \quad \bm{V}_{dq0}^{s}]$ and the tensor product $[\gamma^{r} \quad \omega^{r} \quad \bm{v}^{s}_{dq0}] \otimes [\bm{i}^{s}_{dq0} \quad \bm{I}^{s}_{dq0} \quad \bm{V}^{s}_{dq0}]$. Library $3$ contains all products of $i^{s}_{d}$, $i^{s}_{q}$, $I^{s}_{d}$, $I^{s}_{q}$, $V^{s}_{d}$, and $V^{s}_{q}$. Library $4$ contains second-order polynomials of $\bm{z}$ tensored with first-order goniometric terms of $\gamma^{r}$. Lastly, library $5$ contains third order polynomials of $\bm{z}$, without cubic terms, tensored with first-order goniometric terms of $\gamma^{r}$.

\begin{figure}[t]
    \centering
    \includegraphics[width=0.95\linewidth]{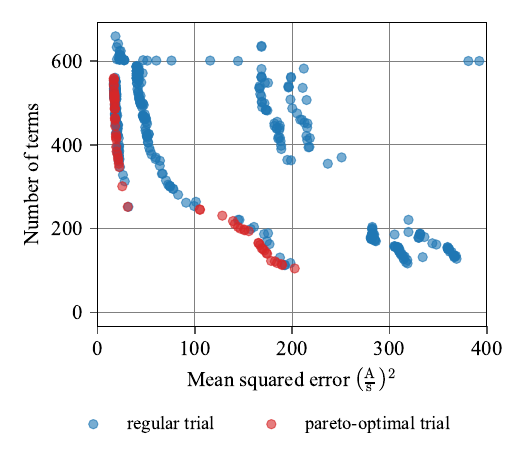}
    \caption{Visualisation of an optimisation study for the stator current dynamical model.}
    \label{fig:pareto}
\end{figure}

For the dynamical model of the stator current and the mappings of the torque and UMP, $500$ trials were evaluated each to identify an appropriate optimiser, its hyperparameters, and library functions. After training, the performance of a trial is evaluated on a validation dataset, which results in a mean-squared error for the given number of active function terms. As visualised in Fig.~\ref{fig:pareto}, a choice is made among Pareto-optimal trials to achieve desired model complexity and performance.

\subsection{Evaluation of Performance and Sparsity}

\begin{figure}[t]
    \centering
    \includegraphics[width=1.0\linewidth]{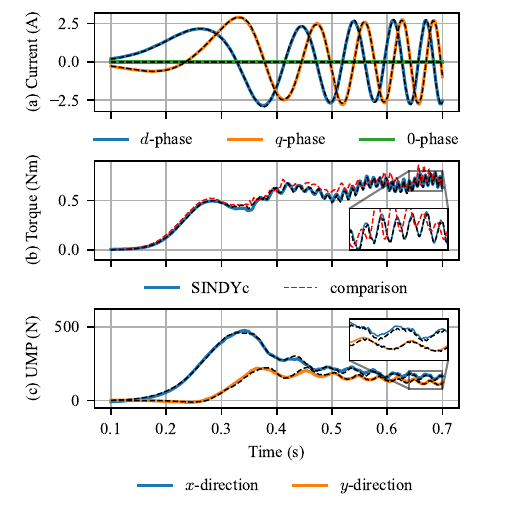}
    \caption{Evaluation of SINDYc model performance for stator current, unbalanced magnetic pull, and electromagnetic torque modelling. The data-generation model is visualised using the black dotted line as a reference.}
    \label{fig:prediction}
\end{figure}

Fig.~\ref{fig:prediction} evaluates SINDYc for the prediction of the stator current, torque, and UMP for a $50\%$ static rotor eccentricity using library $1$, optimised with SR3 with $\nu=10^{-9}$ and $\lambda=1$. Notably, the $0$-phase current remains zero throughout, despite the eccentricity. This is explained by evaluating Kirchhoff's law at the isolated star point of the wye connection, which shows that:
\begin{equation}\label{eq:kirchoff_star}
    i^{s}_{a}+i^{s}_{b}+i^{s}_{c}=0.
\end{equation}
According to (\ref{eq:state_vector_def}), this means that $i^{s}_{0}$ should remain zero, as learned by SINDYc. More importantly, a close correspondence between the $dq$-axes currents and the reference is observed. Although accurate results for the torque and UMP can be obtained through detailed field simulation, this approach is not suitable for control, since one can only rely on measurable quantities or low-cost estimation. For example, the commonly used torque equation \cite{geyer2016model}:
\begin{equation}\label{eq:torque_comp}
    T_{e}=p(\lambda^{s}_{d}i^{s}_{q}-\lambda^{s}_{q}i^{s}_{d}),
\end{equation}
is used in Fig~\ref{fig:prediction}(b) as a comparison. It shows that for eccentric rotors, an improvement can be found using SINDYc. In particular, the effects of magnetic saturation and slotting on the torque are insufficiently captured using (\ref{eq:torque_comp}). However, SINDYc shows only a minor deviation. Although torque estimation can be found in literature, e.g. (\ref{eq:torque_comp}), an estimation of the UMP from measurables eludes the state of the art. In Fig.~\ref{fig:prediction}(c), it is shown that SINDYc can identify a suitable mapping with high accuracy, which could be promising for a data-driven minimisation of the UMP.

\begin{figure}[t]
    \centering
    \includegraphics[width=1.0\linewidth]{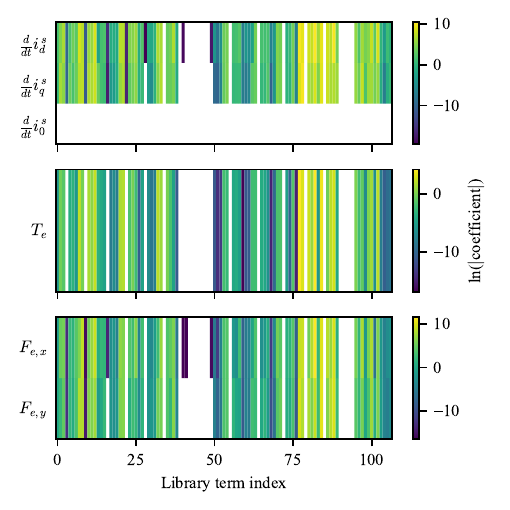}
    \caption{Visualisation of SINDYc weighting matrices. The elements in white are precisely zero.}
    \label{fig:coefficients}
\end{figure}

Fig.~\ref{fig:prediction} shows a close correspondence between SINDYc and reference values. Moreover, SINDYc obtains sparse surrogate models. Fig.~\ref{fig:coefficients} visualises the weighting matrices $\bm{\Xi}$ and $\bm{\Xi}_{\mathcal{M}}$ of the stator current dynamical model and the torque and UMP mappings. Terms that are precisely zero are portrayed in white. The dynamics of $i^{s}_{0}$ are learned using the sparsest solution possible by setting all coefficients equal to $0$, which accurately captures its dynamics as explained by (\ref{eq:kirchoff_star}). Furthermore, SINDYc neglects the usage of library terms with indices $39$ to $49$ and $90$ to $94$. These indices correspond to terms that make use of $i^{s}_{0}$ or $I^{s}_{0}$.

\begin{table}[t]
    \centering
    \caption{Error Evaluation Using Test Datasets}
    \begin{tabular}{|c|c|c|c|}
    \hline
    \textbf{Model type} & \multicolumn{3}{|c|}{\textbf{Mean absolute error (number of terms)}}\\
    \hline
    & \textbf{Centric} & \textbf{Static} & \textbf{Dynamic}\\
    \hline
    $i^{s}_{d}$ (A) & $4.9\cdot10^{-2}$ (33) & $5.3\cdot10^{-2}$ (36) & $5.5\cdot10^{-2}$ (40)\\
    \hline
    $i^{s}_{q}$ (A) & $4.5\cdot10^{-2}$ (34) & $4.7\cdot10^{-2}$ (33) & $5.4\cdot10^{-2}$ (35)\\
    \hline
    $i^{s}_{0}$ (A) & $9.6\cdot10^{-17}$ (0) & $9.6\cdot10^{-17}$ (0) & $9.3\cdot10^{-17}$ (0)\\
    \hline
    $T_{e}$ (Nm) & $1.1\cdot10^{-2}$ (5) & $8.8\cdot10^{-3}$ (39) & $2.7\cdot10^{-2}$ (39)\\
    \hline
    $T_{e,\text{comp}}$ (Nm) & $2.0\cdot10^{-2}$ & $2.5\cdot10^{-2}$ & $3.1\cdot10^{-2}$\\
    \hline
    $F_{e,x}$ (N) & $1.12$ (0) & $4.94$ (56) & $98.96$ (81)\\
    \hline
    $F_{e,y}$ (N) & $1.12$ (0) & $4.80$ (55) & $99.03$ (79)\\
    \hline
    \end{tabular}
    {\footnotesize{\vspace*{5pt}\rule{0pt}{10pt}The torque comparison $T_{e,\text{comp}}$ makes use of (\ref{eq:torque_comp}).}}
    \label{tab:pred_error}
\end{table}

Lastly, the predictive performance of SINDYc is evaluated in the various cases of rotor eccentricity contained in datasets A, B, and C. The results consistently show a low error of the stator current for all eccentricity cases. In contrast, the torque and UMP equations generally increase in complexity and error for the eccentric rotor positions, especially for dynamic eccentricity. SINDYc achieves a low error for centric and statically eccentric rotor positions and thus holds promise for control of the stator current, torque, and UMP. For all eccentricity cases, the torque equations obtained by SINDYc outperform the comparison method of (\ref{eq:torque_comp}). However, the results of dynamic rotor eccentricity illustrate that additional information may be required to execute predictive control in this fault scenario.

\section{Conclusion}
In this work, the sparse regression-based technique, SINDYc, is used to identify models of the stator currents, electromagnetic torque, and UMP for a simulated induction motor with rotor eccentricity. Three datasets are created using a nonlinear magnetic equivalent circuit, namely one with a centric rotor, one with a $50\%$ static rotor eccentricity, and one with a $50\%$ dynamic rotor eccentricity. Explicit knowledge of the stator resistances is avoided by defining stator voltage and current time integrals as new input variables. An identification of the optimiser, its hyperparameters, and the library of candidate functions is handled using a randomised search using $500$ trials per dataset. After evaluation on a test dataset, a Pareto-optimal trial is selected which best achieves the desired balance between model complexity and accuracy.

A dynamical model of the stator current is found that closely approximates the results generated by the magnetic equivalent circuit. Moreover, a low complexity of the stator model is maintained for all rotor eccentricity cases. The mapping of the electromagnetic torque shows an improvement over traditional methods, which could improve torque control performance, especially for drives with unbalances. Interestingly, the results show that the UMP can be estimated with high accuracy in the case of a static rotor eccentricity. Since the data-driven estimation relies solely on measurable quantities, it opens the door for predictive control of the UMP. While this could extend the lifespan of the faulty drive, the creation of datasets for this supervised approach would require measurements of the torque and UMP, e.g., using strain gauges, accelerometers or load cells. In future work, evaluations of SINDYc for the control of electrical drives with unbalances will be explored in simulation and experimentally.

\bibliographystyle{IEEEtran}
\bibliography{refs}

\begin{thebibliography}{10}
\providecommand{\url}[1]{#1}
\csname url@samestyle\endcsname
\providecommand{\newblock}{\relax}
\providecommand{\bibinfo}[2]{#2}
\providecommand{\BIBentrySTDinterwordspacing}{\spaceskip=0pt\relax}
\providecommand{\BIBentryALTinterwordstretchfactor}{4}
\providecommand{\BIBentryALTinterwordspacing}{\spaceskip=\fontdimen2\font plus
\BIBentryALTinterwordstretchfactor\fontdimen3\font minus \fontdimen4\font\relax}
\providecommand{\BIBforeignlanguage}[2]{{%
\expandafter\ifx\csname l@#1\endcsname\relax
\typeout{** WARNING: IEEEtran.bst: No hyphenation pattern has been}%
\typeout{** loaded for the language `#1'. Using the pattern for}%
\typeout{** the default language instead.}%
\else
\language=\csname l@#1\endcsname
\fi
#2}}
\providecommand{\BIBdecl}{\relax}
\BIBdecl

\bibitem{salah2019review}
A.~A. Salah, D.~G. Dorrell, and Y.~Guo, ``A review of the monitoring and damping unbalanced magnetic pull in induction machines due to rotor eccentricity,'' \emph{IEEE Transactions on Industry Applications}, vol.~55, no.~3, pp. 2569--2580, 2019.

\bibitem{geyer2016model}
T.~Geyer, \emph{Model predictive control of high power converters and industrial drives}.\hskip 1em plus 0.5em minus 0.4em\relax John Wiley \& Sons, 2016.

\bibitem{raj2021statespace}
K.~K. Raj, S.~H. Joshi, and R.~Kumar, ``A state-space model for induction machine stator inter-turn fault and its evaluation at low severities by pca,'' in \emph{2021 IEEE Asia-Pacific Conference on Computer Science and Data Engineering (CSDE)}, 2021, pp. 1--6.

\bibitem{brunton2016discovering}
\BIBentryALTinterwordspacing
S.~L. Brunton, J.~L. Proctor, and J.~N. Kutz, ``Discovering governing equations from data by sparse identification of nonlinear dynamical systems,'' \emph{Proceedings of the National Academy of Sciences}, vol. 113, no.~15, pp. 3932--3937, 2016. [Online]. Available: \url{https://www.pnas.org/doi/abs/10.1073/pnas.1517384113}
\BIBentrySTDinterwordspacing

\bibitem{brunton2016sparse}
\BIBentryALTinterwordspacing
------, ``Sparse identification of nonlinear dynamics with control (sindyc),'' \emph{IFAC-PapersOnLine}, vol.~49, no.~18, pp. 710--715, 2016, 10th IFAC Symposium on Nonlinear Control Systems NOLCOS 2016. [Online]. Available: \url{https://www.sciencedirect.com/science/article/pii/S2405896316318298}
\BIBentrySTDinterwordspacing

\bibitem{abdullah2023data}
F.~Abdullah and P.~D. Christofides, ``Data-based modeling and control of nonlinear process systems using sparse identification: An overview of recent results,'' \emph{Computers \& Chemical Engineering}, vol. 174, p. 108247, 2023.

\bibitem{ayankoso2023time}
S.~Ayankoso and P.~Olejnik, ``Time-series machine learning techniques for modeling and identification of mechatronic systems with friction: A review and real application,'' \emph{Electronics}, vol.~12, no.~17, p. 3669, 2023.

\bibitem{dasanayake2024motor}
N.~Dasanayake and S.~Perera, ``Motor state prediction and friction compensation for brushless dc motor drives using data-driven techniques,'' \emph{Nonlinear Dynamics}, pp. 1--16, 2024.

\bibitem{abu2023data}
M.~A. Abu-Seif, M.~Ahmed, M.~Y. Metwly, A.~S. Abdel-Khalik, M.~S. Hamad, S.~Ahmed, and N.~Elmalhy, ``Data-driven-based vector space decomposition modeling of multiphase induction machines,'' \emph{IEEE Transactions on Energy Conversion}, vol.~38, no.~3, pp. 2061--2074, 2023.

\bibitem{desenfans2024influence}
P.~Desenfans, Z.~Gong, D.~Vanoost, K.~Gryllias, J.~Boydens, and D.~Pissoort, ``The influence of the unbalanced magnetic pull on fault-induced rotor eccentricity in induction motors,'' \emph{Journal of Vibration and Control}, vol.~30, no. 5-6, pp. 943--959, 2024.

\end{thebibliography}

\end{document}